\documentclass[final,5p,times]{elsarticle}
 \usepackage{graphics}
\usepackage{amssymb}
\pdfoutput=1
 \usepackage{lineno}

\journal{NIM A  RICAP-2013}

\begin{document}

\begin{frontmatter}



\title{ The prototyping/early construction phase of the BAIKAL-GVD project }


\author[i1]{A.D. Avrorin}
\author[i1]{A.V. Avrorin}
\author[i1]{V.M. Aynutdinov}
\author[i7]{R. Bannasch}
\author[i2]{I.A. Belolaptikov}
\author[i3]{D.Yu. Bogorodsky}
\author[i2]{V.B. Brudanin}
\author[i3]{N.M. Budnev}
\author[i1]{I.A. Danilchenko}
\author[i1]{G.V. Domogatsky}
\author[i1]{A.A. Doroshenko}
\author[i2]{A.N. Dyachok}
\author[i1]{Zh-A.M. Dzhilkibaev}
\author[i5]{S.V. Fialkovsky}
\author[i3]{A.R. Gafarov}
\author[i1]{O.N. Gaponenko}
\author[i1]{K.V. Golubkov}
\author[i3]{T.I. Gress}
\author[i2]{Z. Honz}
\author[i7]{K.G. Kebkal}
\author[i7]{O.G. Kebkal}
\author[i2]{K.V. Konishchev}
\author[i3]{E.N. Konstantinov}
\author[i3]{A.V. Korobchenko}
\author[i1]{A.P. Koshechkin}
\author[i1]{F.K. Koshel}
\author[i4]{V.A. Kozhin}
\author[i5]{V.F. Kulepov}
\author[i1]{D.A. Kuleshov}
\author[i1]{V.I. Ljashuk}
\author[i3]{A.I. Lolenko}
\author[i5]{M.B. Milenin}
\author[i3]{R.A. Mirgazov}
\author[i4]{E.A. Osipova}
\author[i1]{A.I. Panfilov}
\author[i3]{L.V. Pan'kov}
\author[i3]{A.A. Perevalov}
\author[i2]{E.N. Pliskovsky}
\author[i3]{V.A. Poleshuk}
\author[i6]{M.I. Rozanov}
\author[i3]{V.F. Rubtsov}
\author[i2]{E.V. Rjabov}
\author[i2]{B.A. Shaybonov}
\author[i1]{A.A. Sheifler}
\author[i4]{A.V. Skurikhin}
\author[i2]{A.A. Smagina}
\author[i1]{O.V. Suvorova}
\author[i3]{B.A. Tarashchansky}
\author[i7]{S.A. Yakovlev}
\author[i3]{A.V. Zagorodnikov}
\author[i1]{V.A. Zhukov}
\author[i3]{V.L. Zurbanov}

\address[i1]{Institute for Nuclear Research, Moscow, 117312 Russia}
\address[i2]{Joint Institute for Nuclear Research, Dubna, 141980 Russia}
\address[i3]{Irkutsk State University, Irkutsk, 664003 Russia}
\address[i4]{Institute of Nuclear Physics, Moscow State University, Moscow, 119991 Russia}
\address[i5]{Nizhni Novgorod State Technical University, Nizhni Novgorod, 603950 Russia}
\address[i6]{St. Petersburg State Marine Technical University, St. Petersburg, 190008 Russia}
\address[i7]{EvoLogics GmbH, Berlin, Germany}
%\address[i9]{Russian Research Center Kurchatov Institute, Moscow, 123182 Russia}

\begin{abstract}
The Prototyping phase of the BAIKAL-GVD project has been started in April 2011 with the deployment of a three string
engineering array which comprises all basic elements and systems of the Gigaton Volume Detector (GVD) in Lake Baikal. In
April 2012 the version of engineering array which comprises the first full-scale string of the GVD demonstration cluster has
been deployed and operated during 2012. The first stage of the GVD
demonstration cluster which consists of three strings is deployed in
April 2013. We review the Prototyping phase of the BAIKAL-GVD project and describe the configuration and design of the
2013 engineering array.
\end{abstract}

\begin{keyword}
neutrino \sep neutrino telescope \sep Baikal
\end{keyword}

\end{frontmatter}

% \linenumbers

\section{Introduction}
\label{s1}
 The BAIKAL-GVD Project is a logical extension
 of the research and development work performed
 over the last several years by the BAIKAL
 Collaboration. The optical properties of the lake
 deep water have been established \cite{water}, and the
 detection of high-energy neutrinos has been
 demonstrated with the existing detector NT200 \cite{HE2006,HE2009}.
 This achievement represents a proof of concept
 for commissioning new instrument, Gigaton Volume
 Detector (BAIKAL-GVD), with superior detector
 performance and an effective telescope size at or
 above the kilometer-scale.
 The next generation neutrino telescope BAIKAL-GVD 
in Lake Baikal will be research infrastructure
 aimed primarily at studying astrophysical neutrino
 fluxes and particularly, mapping the high-energy
 neutrino sky in the Southern Hemisphere including
 the region of the galactic center. Other topics include
 indirect search for dark matter by detecting neutrinos
 produced in WIMPs annihilation in the Sun or in the
 center of the Earth. GVD will also search for exotic
 particles like magnetic monopoles, super-symmetric
 Q-balls or nuclearites. The detector will utilize Lake
 Baikal water instrumented at depth with light sensors
 that detect the Cherenkov radiation from secondary
 particles produced in interactions of high-energy
 neutrinos inside or near the instrumented water
 volume. Signal events consist of up-going muons
 produced in neutrino interactions in the bedrock or
 the water, as well as of electromagnetic and hadronic
 showers (cascades) from CC-interactions of 
$\nu_e$ and $\nu_{\tau}$
 or NC-interactions of all flavors inside the array
 detection volume. Background events are mainly
 downward-going muons from cosmic ray
 interactions in the atmosphere above the detector.
 The site chosen for the experiment is in the
 southern basin of Lake Baikal, near the outfall of a
 small river, named Ivanovka, and about 40 km west
 of the place, where the Angara river leaves the lake.
 Here combination of hydrological, hydrophysical,
 and landscape factors is optimal for deployment and
 operation of the neutrino telescope. The
 geographical coordinates of the detector site are
51$^{\circ}$50$^{\prime}$N and 104$^{\circ}$20$^{\prime}$E.
 Lake depth is about 1360 m here at distances
 beginning from about of three kilometers from the
 shore. A flat the lake bed throughout several tens of
 kilometers from the shore allows practically
 unlimited instrumented water volume for deep
 underwater Cherenkov detector. A strong up to 1 m
 thick ice cover from February to the middle of April
 allows telescope deployment, as well as maintenance
 and research works directly from the ice surface,
 using it like a solid and fixed assembling platform.
 The quality of the ice cover, the absence of stable
 hummocking fields and backbone slits determining conditions from the viewpoint of safety
 equipment assembling and underwater cable lines
 deployment. The period of safety works is usually
 longer than 8 weeks. Vanishingly small values of under-ice water currents allow the required precision
 of the assembling works.
 The light propagation in the Baikal water characterized by an absorption length of about 20--25
 m and a scattering length of 30--50 m. The water
 luminescence is moderate at the detector site. The
 rate of light pulses from $^{40}$K-decays is negligible.
 The first generation Baikal Neutrino Telescope
 NT200 is operating in Lake Baikal since April 1998
 \cite{AP97,AP08,AL11}. The upgraded Baikal telescope NT200+
 was commissioned in April, 2005, and consists of central part (the former, densely instrumented
 NT200 telescope) and three additional external
 strings. The deployment of the NT200+ was a first
 step towards a km$^3$-scale neutrino telescope in Lake
 Baikal.
 The first prototype of the GVD electronics was
 installed in Lake Baikal in April 2008 \cite{NIM602}. It was reduced-size section with 6 optical modules (OMs).
 This detection unit provided the possibility to study
 basic elements of the future detector: new optical
 modules and FADC based measuring system. During
 the next two years different versions of prototype
 string were tested in Lake Baikal as a part of the NT200+ detector. The 2009 prototype string consists
 of 12 optical modules with six photomultiplier tubes
 (PMTs) R8055 and six XP1807 \cite{NIM626,NIM630}. In April 2010,
 the string with 8 PMTs R7081HQE and 4 PMTs
 R8055 was deployed in Lake Baikal. The operation
 of these prototype strings in 2009 and 2010 allows first assessment
 of the DAQ performance \cite{NIM639,PTE54,NIM692}.
%%%%%%%%%%%%%%%%%%%%%%%%%%%%%%%%%%%%%%%%%%%%%%%%%%%%%%%%%%%
% \begin{figure*}[!t]
 \begin{figure}[t]
  \centering
  \includegraphics[width=0.4\textwidth]{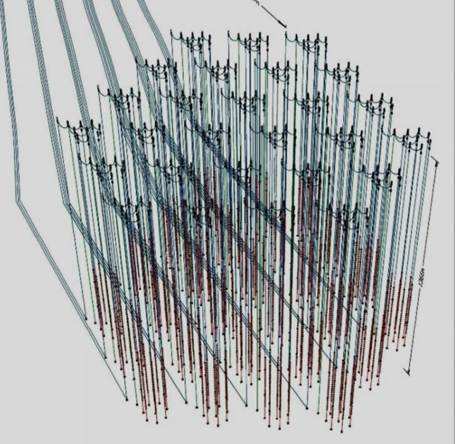}
  \caption{Artistic view of the GVD-telescope.}
  \label{fig1}
% \end{figure*}
 \end{figure}
%%%%%%%%%%%%%%%%%%%%%%%%%%%%
 \begin{figure}[t]
  \centering
  \includegraphics[width=0.4\textwidth]{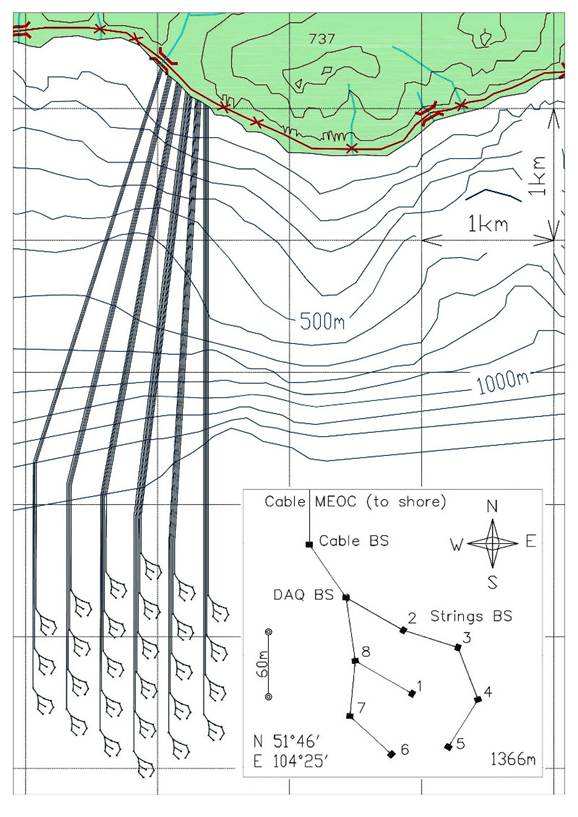}
  \caption{Layout of the GVD. In inner box layout of the GVD-cluster is
    shown.}
  \label{fig2}
 \end{figure}
%%%%%%%%%%%%%%%%%%%%%%%%%%%%%%%%%%%%%%%%%%%%%%%%

\section{GVD design}
\label{s2}
 The concept of BAIKAL-GVD is based on a
 number of evident requirements to the design and
 architecture of the recording system of the new
 array: the utmost use of the advantages of array
 deployment from the ice cover of Lake Baikal, the
 extendability of the facility and provision of
 effective operation even in the first stage of
 deployment, and the possibility implementing
 different versions of arrangement and spatial
 distribution of light sensors within the same
 measuring system.

 The design for the BAIKAL-GVD neutrino
 telescope is an array of 10386 photomultiplier tubes
 each enclosed in a transparent pressure sphere to
 comprise an optical module\footnote{Design of array with 2034 OMs was discussed earlier
elsewhere \cite{NIM630,NIM639,NIM692}.}. The OMs are arranged
 on vertical load-carrying cables to form strings. The
 basic configuration of telescope consists of 27  clusters of strings - functionally independent
 subarrays, which are connected to shore by
 individual electro-optical cables (see figure 1, 2).
 Each cluster comprises eight 705 m long strings of
 optical modules -- seven peripheral strings are
 uniformly arranged at a 60 m distance around a
 central one. Each string comprises 48 OMs spaced
 by 15 m at depths of 600 m to 1300 m below the
 surface. All OMs are faced downward. OMs on each
 string are combined in four sections -- detection units
 of telescope. The distances between the central
 strings of neighboring clusters are H=300 m. The
 clusters are spaced over an area of approximately 2
 km$^2$. The water volume instrumented by OMs is
 about of 1.4 km$^3$.

The objective of the optimization of the GVD
design was to provide a large cascade detection
volume with the condition of also effective recording
 high energy muons. Muon effective areas for two
 optimized GVD configurations are shown in figure
 3. The curves labeled by GVD*4 and GVD relate to
 configurations with 10368 OMs and 2304 OMs,
 respectively. Muon effective area (6/3 event
 selection requirement - at least 6 hit channels on at
 least 3 strings) rises from 0.3 km$^2$ at 1 TeV to about
 of 1.8 km$^2$ asymptotically. The fraction of events induced
by muons (E$_{\mu}>$ 1 TeV)
 with mismatch angles between generated and
 reconstructed muon directions less than a given
 value $\psi$ is shown in figure 4. Muon arrival direction
 resolution (median mismatch angle) is about of 0.25
 degree.

 Shower effective volumes for two GVD
 configurations are shown in figure 5. Shower
 effective volumes (11/3 condition - at least 11 hit
 channels on at least 3 strings) for basic configuration
 are about of 0.4--2.4 km$^3$ above 10 TeV. The
 accuracy of shower energy reconstruction is about of
 20--35\% depending on shower energy. The accuracy  of a shower
 direction reconstruction is about of 
3.5--6.5 degrees (median value). Distribution of the
 mismatch angle between generated and
 reconstructed directions of 1 PeV showers is shown
 in figure 6.

%%%%%%%%%%%%%%%%%%%%%%%%%%%%%%%%%%%%%%%%%%%%%%%%%%%%%%%%%%%%%%%%%%%%%%%%%%%%%%
%%%%%%%%%%%%%%%%%%%%%%%%%%%%%%%%%%%%%%%%%%%%%%%%%%%%%%%%%%%
% \begin{figure*}[!t]
 \begin{figure}[t]
  \centering
  \includegraphics[width=0.4\textwidth]{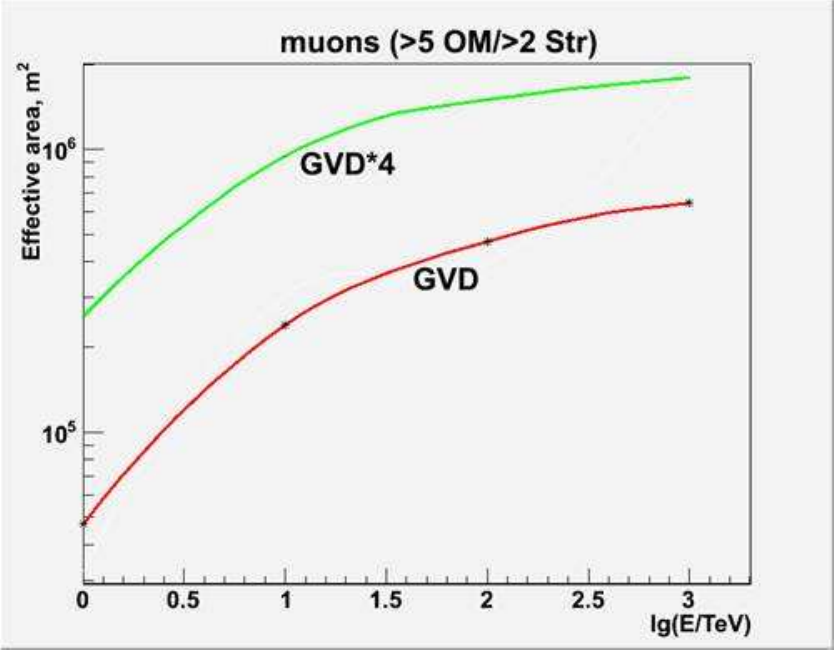}
  \caption{Muon effective area. The curves labeled by GVD*4 and GVD
    relate to configurations with 10368 OMs and 2304 OMs, respectively.}
  \label{fig3}
% \end{figure*}
 \end{figure}
%%%%%%%%%%%%%%%%%%%%%%%%%%%%
%%%%%%%%%%%%%%%%%%%%%%%%%%%%%%%%%%%%%%%%%%%%%%%%%%%%%%%%%%%
% \begin{figure*}[!t]
 \begin{figure}[t]
  \centering
  \includegraphics[width=0.4\textwidth]{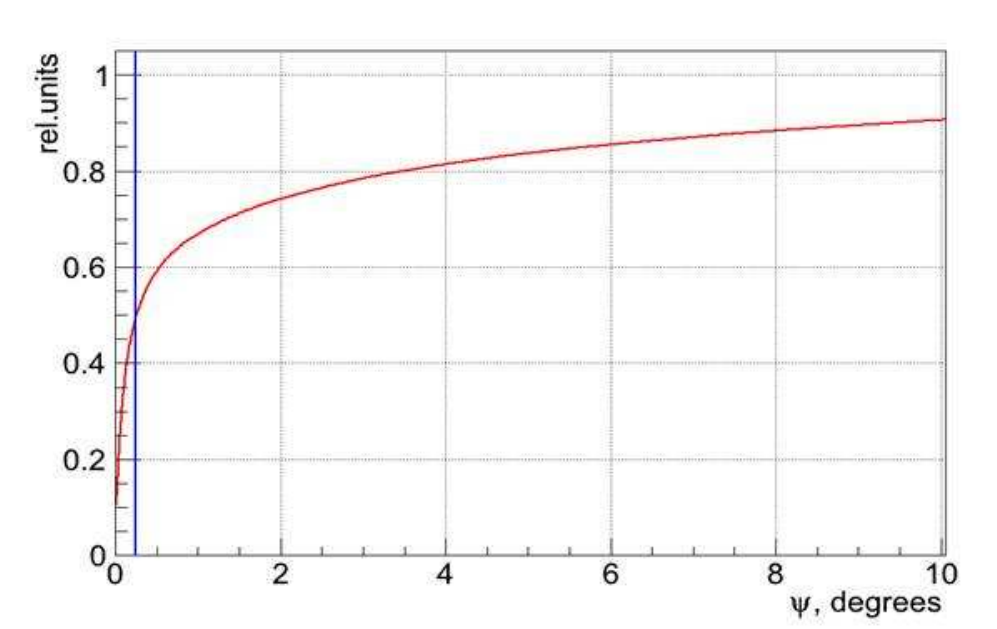}
  \caption{The fraction of muon events (E$_{\mu}>1$ TeV) with mismatch angle $\psi$ less than
    a given value.}
  \label{fig4}
% \end{figure*}
 \end{figure}

%%%%%%%%%%%%%%%%%%%%%%%%%%%%

 \begin{figure}[t]
  \centering
  \includegraphics[width=0.4\textwidth]{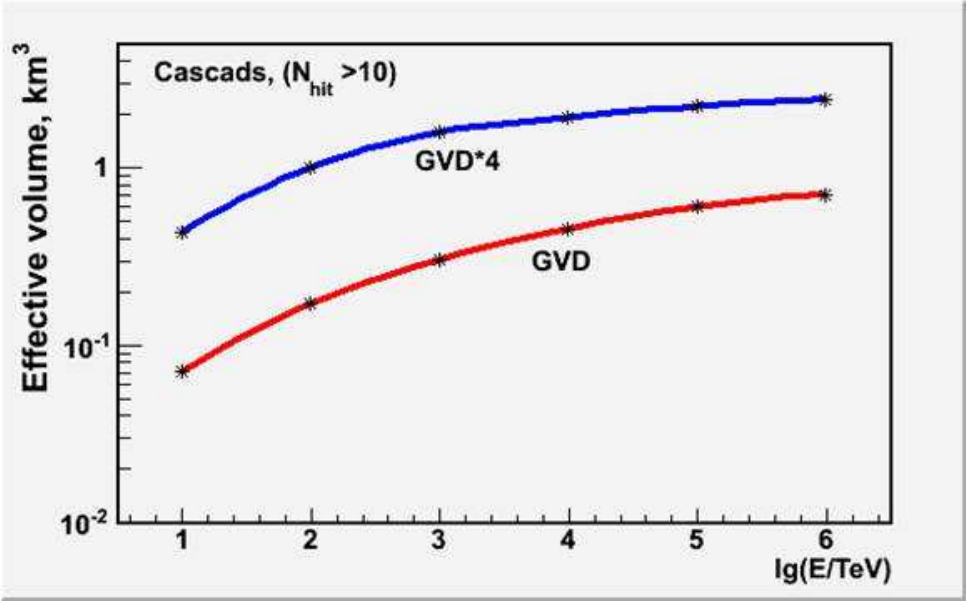}
  \caption{Effective volume of cascades detection. The
curves labeled by GVD*4 and GVD relate to configurations with 10368 OMs and 2304
OMs, respectively.}
  \label{fig5}
 \end{figure}
%%%%%%%%%%%%%%%%%%%%%%%%%%%%%%%%%%%%%%%%%%%%%%%%

%%%%%%%%%%%%%%%%%%%%%%%%%%%%%%%%%%%%%%%%%%%%%%%%%%%%%%%%%%%
% \begin{figure*}[!t]
 \begin{figure}[t]
  \centering
  \includegraphics[width=0.4\textwidth]{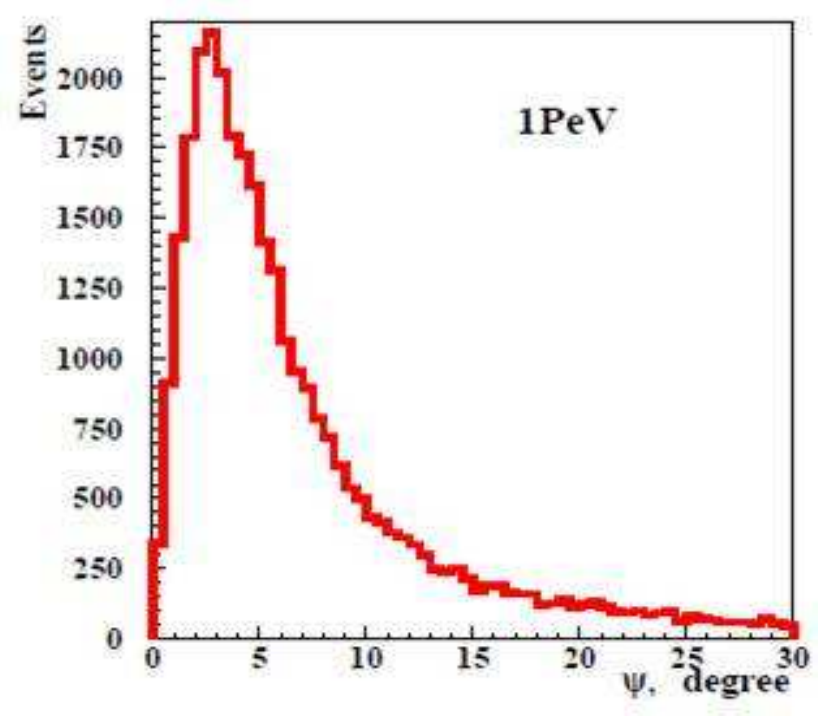}
  \caption{Distribution of the mismatch angle $\psi$ between generated
and reconstructed 1 PeV shower directions.
}
  \label{fig6}
% \end{figure*}
 \end{figure}

%%%%%%%%%%%%%%%%%%%%%%%%%%%%

%%%%%%%%%%%%%%%%%%%%%%%%%%%%%%%%%%%%%%%%%%%%%%%%%%%%%%%%%%%
% \begin{figure*}[!t]
 \begin{figure}[t]
  \centering
  \includegraphics[width=0.4\textwidth]{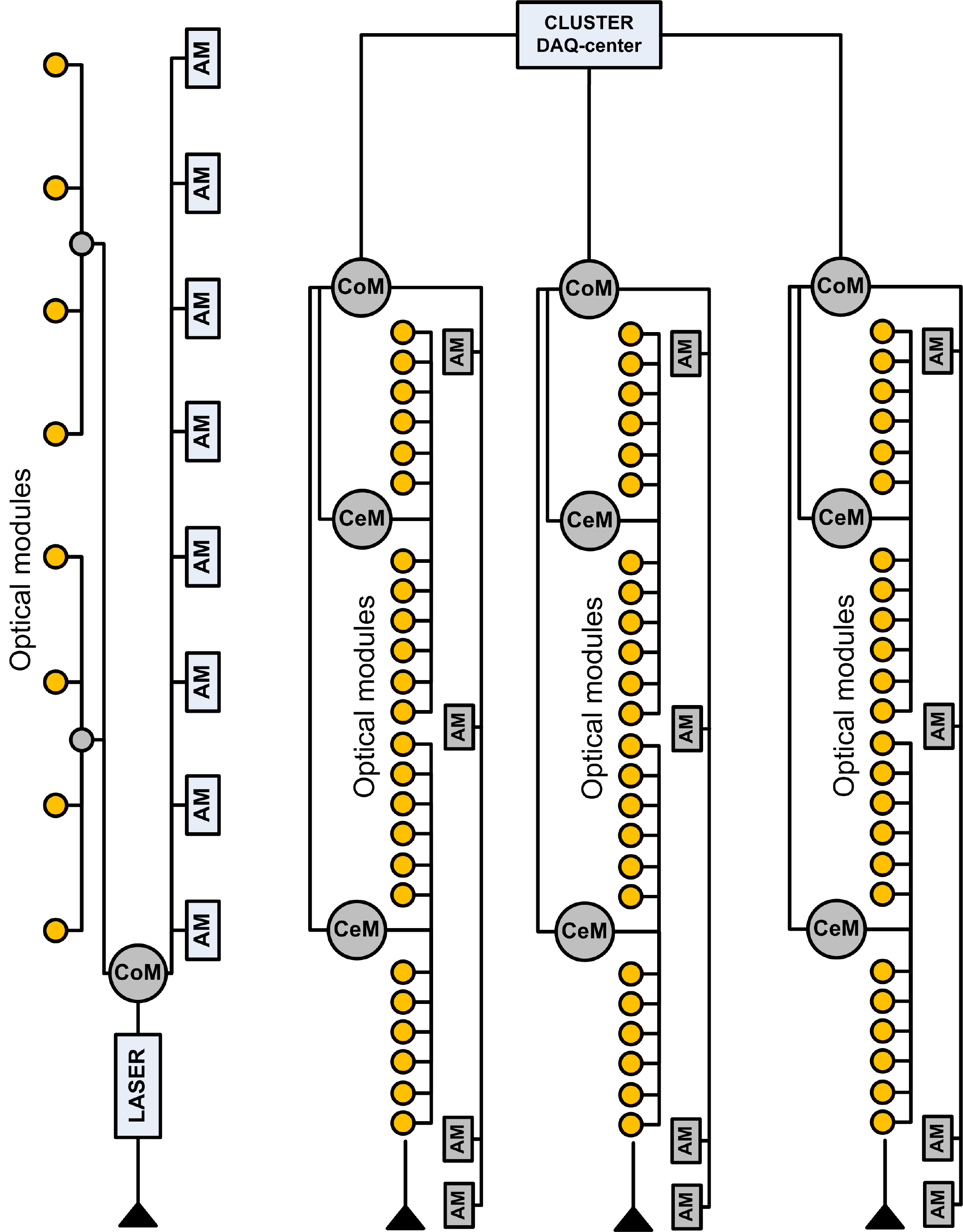}
  \caption{Schematic drawing of the 2013 year engineering array.}
  \label{fig7}
% \end{figure*}
 \end{figure}

%%%%%%%%%%%%%%%%%%%%%%%%%%%%

\section{Prototype arrays}
\label{s3}
 The prototyping phase of the BAIKAL-GVD
 project aims at in situ comprehensive tests of all
 elements and systems of the future telescope as the
 parts of engineering arrays operating in Lake Baikal.
 Prototyping phase will be concluded with
 deployment in 2015 of the first demonstration cluster
 of the GVD in Lake Baikal. Demonstration cluster
 will comprise total of 192 optical modules arranged
 on eight 345 m long strings (7 side strings located at
 60 m distances from a central one). Each string
 comprises 24 OMs spaced by 15 m at depths of 950
 m to 1300 m below the surface. OMs on each string
 are combined in two sections. Also the
 demonstration cluster will comprise an acoustic
 positioning system and instrumentation string with
 equipment for array calibration and monitoring of
 environment parameters.

 In April 2011 the first autonomous engineering
 array which includes preproduction modules of all
 elements, measuring and communication systems, as
 well as prototype of acoustic positioning system of
 GVD-cluster has been installed and commissioned in
 Lake Baikal \cite{NIM692}. Array comprises total of 24 optical
 modules with the different types of PMTs (16 PMTs
 R7081HQE, 3 PMTs XP1807 and 5 PMTs R8055)
 which are arranged on three 70 m long vertical
 strings. Distances between strings are about of 40 m.
 Eight OMs on each string are spaced by 10 m and
 form one section.

 In April 2012 the next version of engineering
 array which comprises 36 OMs has been deployed in
 Lake Baikal. This array consists of two short and one
 long string. Each of short strings consists of six OMs
 which are combined in one section. The long string
 comprises 24 OMs with R7081HQE PMTs
 combined in two sections. Vertical spacing of OMs
 is 15 m. This string is the first full-scale string of the
 GVD demonstration cluster.

 The next important step on realization of the
 GVD project was made in 2013 by deployment of
 enlarged engineering array which comprises 72 OMs
 arranged on three 345 m long full-scale strings of the
 GVD demonstration cluster, as well as
 instrumentation string with an array calibration and
 environment monitoring equipment. The artistic
 view of this engineering array is shown in figure 7.
 The vertical spacing of OMs is 15 m and the
 horizontal distance between strings is about of 40 m.
 In addition to OMs each string comprises the
 communication module (CoM), and two central
 modules of the sections (CeM), as well as one
 transmitter and 3 receivers of acoustic positioning
 system (AM) \cite{PTE56}. The modified cluster DAQ-center
 is located at separate cable station and is connected
 to shore by electro-optical cable.

%%%%%%%%%%%%%%%%%%%%%%%%%%%%%%%%%%%%%%%%%%%%%%%%%%%%%

\section{ Data acquisition}
\label{s4}
 The Data Acquisition System of the engineering
 array is formed from three basic units: optical
 modules, sections of OMs, and cluster of the
 sections. 

Each OM (figure 8) contains a
 photomultiplier tube Hamamatsu R7081HQE,
 which detects the Cherenkov light produced by
 relativistic charged particles passing through the
 water. Electronics of optical module is discussed in
 details elsewhere \cite{NIM602,NIM626,NIM630,NIM639}. Latest version of OM
 design differs from previous ones by a power supply
 scheme and a type of used deep underwater
 connectors. The coaxial single-contact connectors
 which were used earlier were replaced by SubConn
 five-contact connectors. These modernizations allow
 an improvement of reliability of the underwater
 recording system and employment of different lines
 of OM cable for PMT signal transmission and OM
 power supply. Moreover it allows replacement of a
 low noisy OM power sources with relatively low
 efficiency (~60\%) by a switching power supply with
 efficiency about of 90\%. The performed
 modifications of OMs have decreased a power
 consumption of array on more than 20\%.

%%%%%%%%%%%%%%%%%%%%%%%%%%%%%%%%%%%%%%%%%%%%%%%%%%%%%%%%%%%
% \begin{figure*}[!t]
 \begin{figure}[t]
  \centering
  \includegraphics[width=0.4\textwidth]{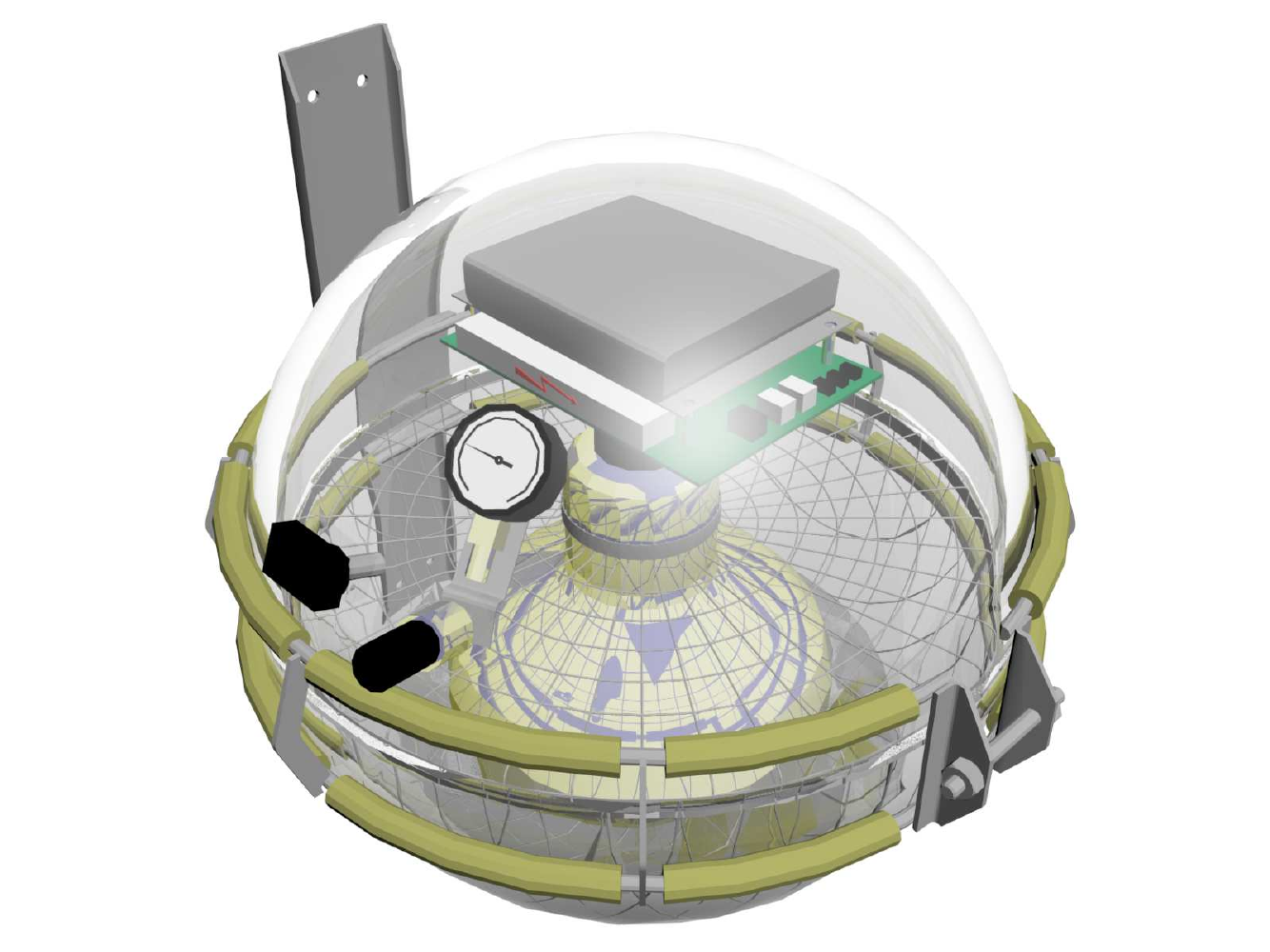}
  \caption{Optical module.
}
  \label{fig8}
% \end{figure*}
 \end{figure}
%%%%%%%%%%%%%%%%%%%%%%%%%%%%

 The optical modules are grouped into sections,
 the detection units of array \cite{NIM602,NIM626,NIM630,NIM639,PTE54}. Each section
 includes 12 OMs and the central module. PMT’s
 signals from all OMs are transmitted through 90 m
 long coaxial cables to the central module of the
 section, where they are digitized by custom-made
 ADC boards with 200 MHz sampling rate. The
 waveform information from all measuring channels
 of the section is transferred to the Master board
 located in the CeM. The Master board provides data
 readout from ADC, connection via local Ethernet to
 the cluster DAQ-center, control of the section
 operation and the section trigger logic \cite{NIM602,NIM626,NIM630,NIM639,PTE54}.
 Request of the section trigger is transferred from the
 Master board to the cluster DAQ-center, where a
 global trigger for all sections is formed. The global
 trigger initiates data transmission from all sections to
 shore.

 Data collected in section central models are
 transmitted to shore through three different segments
 of underwater communication network based on
 Ethernet. The section communication channels
 connect each CeM with the corresponding CoM.
 Given the lengths of these communication channels
 more than 100 m, the shDSL modems are used as the
 Ethernet extensions for data transmission from CeM
 to CoM. In CoM the section communication
 channels are joined into a single one, which connects
 each section with the cluster DAQ-center. Data
 transmission between each CoM and the cluster
 DAQ-center are also based on shDSL technology.

 The data transmission between the cluster DAQ-center and shore station is provided through optical
 fibre lines extended at about of 6 km. Maximal speed
 of data transmission to shore is limited by band
 width of a connection channel between a string CoM
 and the cluster DAQ-center and is about of 8 Mbit/s.
 To provide the required data rate (not less than 100
 Hz), online data processing in each section is
 performed. As a result a raw data sample is reduced
 more than 50 times, since the data are refined by the
 Master electronic cards located in CeMs. To provide
 required speed of data processing the ADC and
 Master cards were upgraded in 2013 with
 replacement of the FPGA Spartan 3 by the Spartan 6
 microcircuits.

%%%%%%%%%%%%%%%%%%%%%%%%%%%%%%%%%%%%%%%%%%%%%%%%%%
\section{ Slow control}
\label{s5}
 Basic functions of control and monitoring of
 array recording system are realized by OMs
 controllers and ADC and Master boards. It means a
 setup of array operation mode, setting the thresholds
 and the data accumulating time ranges of measuring
 channels, PMT's gains, control and monitoring of
 equipment parameters and background conditions
 during array operation. As an example, in figure 9
 are shown results of long-term measurements of the
 channels counting rates of third section of the 2013
 array. The main contribution to recorded counting
 rates is caused by a chemiluminescence of the deep
 water. During April--June 2013 there were not
 observed bright bursts provided by water luminosity.

%%%%%%%%%%%%%%%%%%%%%%%%%%%%%%%%%%%%%%%%%%%%%%%%%%%%%%%%%%%
% \begin{figure*}[!t]
 \begin{figure}[t]
  \centering
  \includegraphics[width=0.4\textwidth]{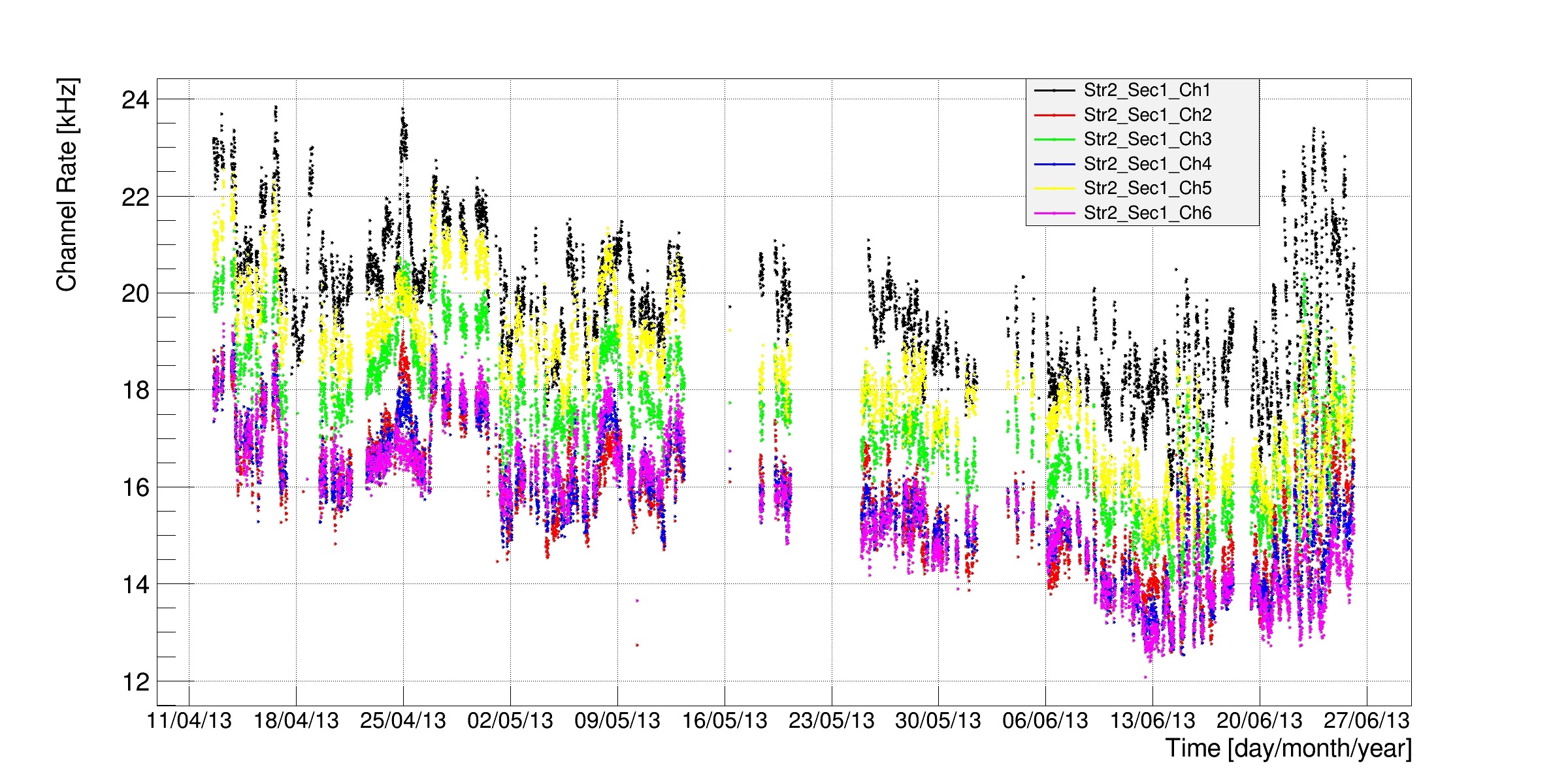}
  \caption{Counting rates of OMs 1-6 belonging to the bottom
 section of second string which were observed in April-June 2013.
 Numbering of channels is directed from the bottom to top of the section.
}
  \label{fig9}
% \end{figure*}
 \end{figure}
%%%%%%%%%%%%%%%%%%%%%%%%%%%%

 In 2013 electrical power system of engineering
 array was upgraded substantially. Instead of scheme
 with disconnecting switches of 300 VDC power
 lines which were installed in underwater modules
 and were managed through power supply lines, 
12-channel 300 VDC commutators controlled via
 Ethernet are used now. They were developed and
 successfully tested in 2012. This modification
 remarkably simplifies a power control and improves
 the array reliability in whole.
 Schematic view of 300 VDC commutation
 is shown in figure 10. The independent switching of
 the commutator channels is controlled by COM-
 server and 16-channel digital output module ICP
 DAS I-7045. Monitoring of output voltage is
 processed by 20-channel analog input module ICP
 DAS I-7017Z.

 Cluster power supply system is divided in two
 levels (see figure 11). First level includes the 
12-channel commutator which is used for string power
 supply and is placed in the cluster DAQ-center.
 Second level is formed by the commutators of strings
 which are located inside the string communication
 modules (CoM). They provide independent switch
 on/off of power supply of the string sections. This
 scheme of power supply provides a confident
 operation of the array in whole even if some
 elements of recording system like section or string
 would be broken.

OMs power supply management is not changed
in 2013: switch on/off of OM voltage (12 VDC) in
each section is processed by 12-channel relay board
which is controlled by Master board trough RS485
bus.

%%%%%%%%%%%%%%%%%%%%%%%%%%%%%%%%%%%%%%%%%%%%%%%%%%%%%%%%%%%
% \begin{figure*}[!t]
 \begin{figure}[t]
  \centering
  \includegraphics[width=0.4\textwidth]{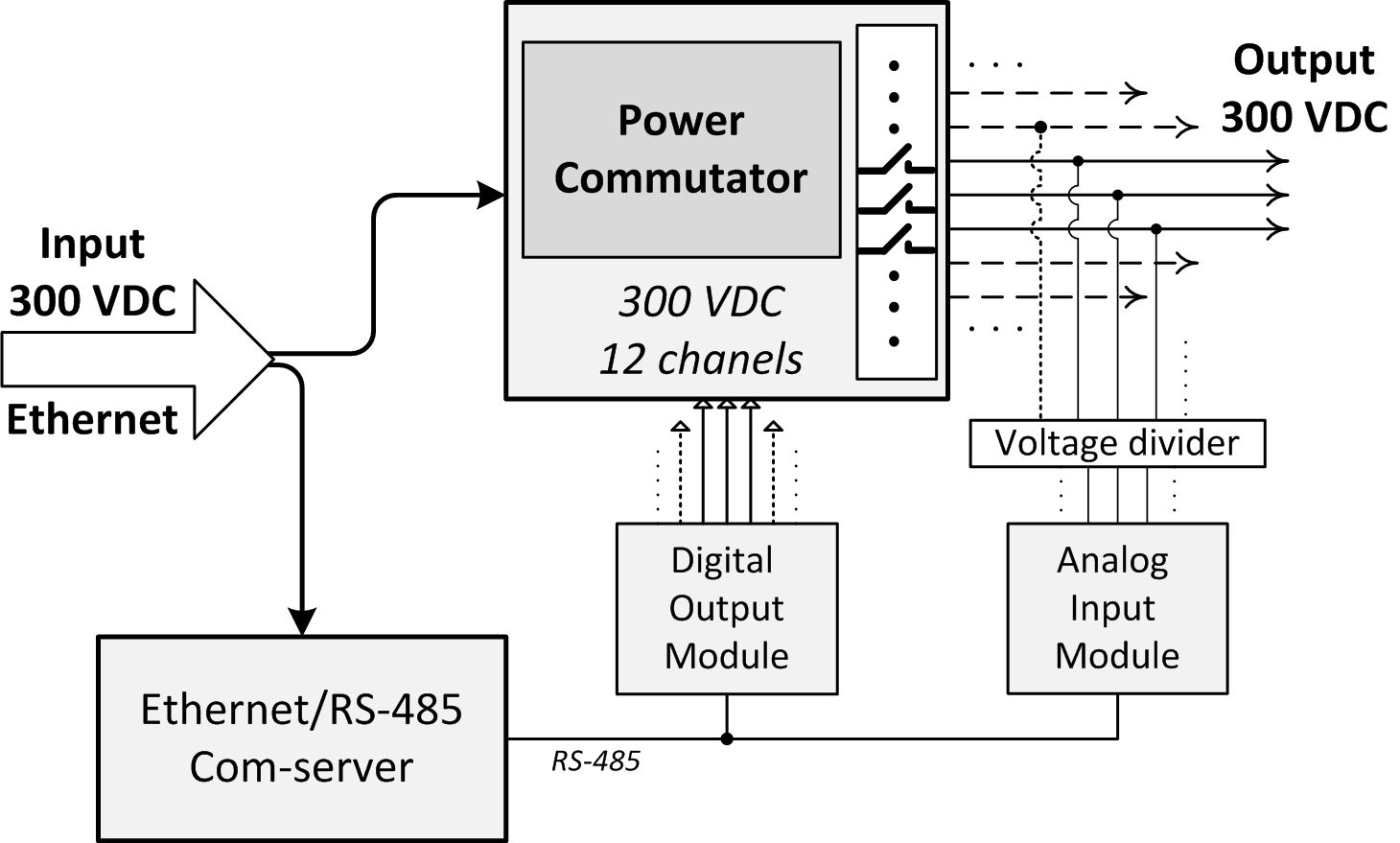}
  \caption{Schematic view of 300 VDC commutation.
}
  \label{fig10}
% \end{figure*}
 \end{figure}
%%%%%%%%%%%%%%%%%%%%%%%%%%%%
%%%%%%%%%%%%%%%%%%%%%%%%%%%%%%%%%%%%%%%%%%%%%%%%%%%%%%%%%%%
% \begin{figure*}[!t]
 \begin{figure}[t]
  \centering
  \includegraphics[width=0.4\textwidth]{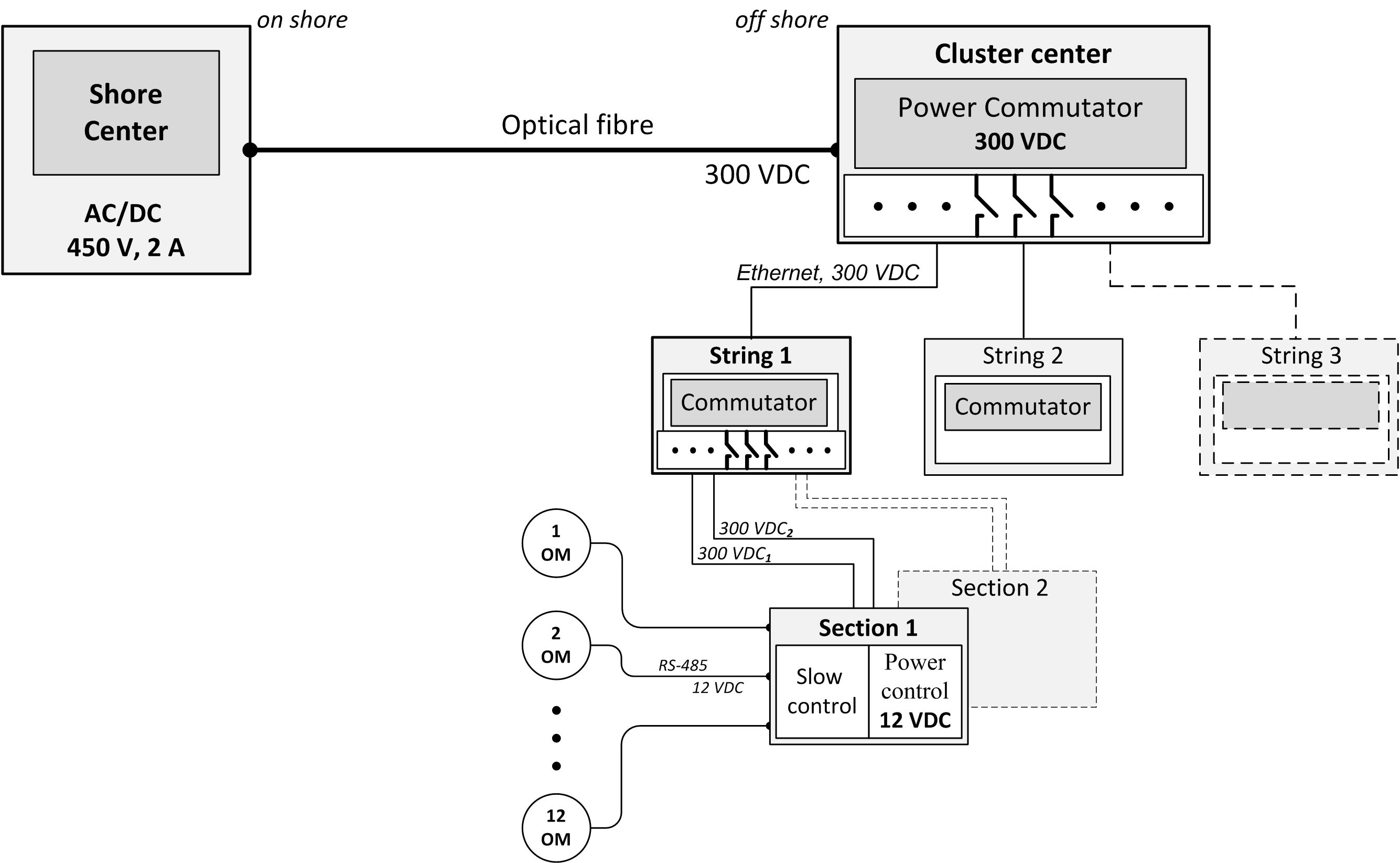}
  \caption{Schematic view of a cluster power supply architecture.
}
  \label{fig11}
% \end{figure*}
 \end{figure}
%%%%%%%%%%%%%%%%%%%%%%%%%%%%
%%%%%%%%%%%%%%%%%%%%%%%%%%%%%%%%%%%%%%%%%%%
\section{Calibration}
\label{s6}
 Calibration of the array recording system consists
 of the following procedures: amplitude and temporal
 calibrations of the measuring channels and time
 calibration of the sections. All these calibration
 procedures are based on usage of OM's internal
 calibration LEDs (two LEDs inside each OM). The
 LED’s intensities as well as time delays between
their light pulses are selected depending on the used
calibration mode.

 We apply a standard procedure of the
 amplitude calibration of PMTs based on an analysis
 of a single photoelectron spectrum (s.p.e.). In this
 calibration mode the pulses of two LEDs of OM are
 used. Intensity of the first LED is fitted to provide a
 detection of s.p.e. signals with detection probability
 about of 10\%. These pulses are used to measure
 s.p.e. distribution of channel signals. Pulses of the
 second LED with intensities corresponding to about
 of 50 p.e. PMT's signal are delayed on 500 ns and
 are used as a trigger to suppress background signals
 with small amplitudes initiated by PMT dark current,
 as well as light background of the lake deep water.

 During the temporal calibration of measuring
 channels the relative offsets of signals recorded by
 OMs are derived. Temporal signal delay of
 measuring channel is formed by PMT’s internal
 delay and delay caused by signal passing through
 about of 90 m long cable connecting OM and SeM.
 Cable delays are measured once in the laboratory
 and are the same during array operation. PMT delay
 depends on a power voltage and thus it requires
 regular calibration during array operation. There is a
 specialized test pulse which is generated by OM
 controller and is delivered to point of signal creation
 in PMT preamplifier. Test pulse initiation is
 synchronized with start time of LED. From
 measured difference between arrival times of LED
 signal and test pulse the PMT delay is obtained.

 Intensities of LEDs light bursts are high enough
 to be detected by PMTs of neighboring sections and
 strings. It allows synchronization between OMs of
 different sections and different strings, as well as to
 check out a precision of found delays between
 channels inside one section. The calibration
 coefficients are defined on a base of known positions
 of strings and OMs obtained from analysis of data
 accumulated by acoustic positioning system. In this
 calibration mode pairs of light pulses delayed by 500
 ns are used. This provides a reliable way to select the
 calibration signals from PMT's background.

%%%%%%%%%%%%%%%%%%%%%%%%%%%%%%%%%%%%%%%%%%%%%%%%%%%%%%%%%%%
% \begin{figure*}[!t]
 \begin{figure}[t]
  \centering
  \includegraphics[width=0.4\textwidth]{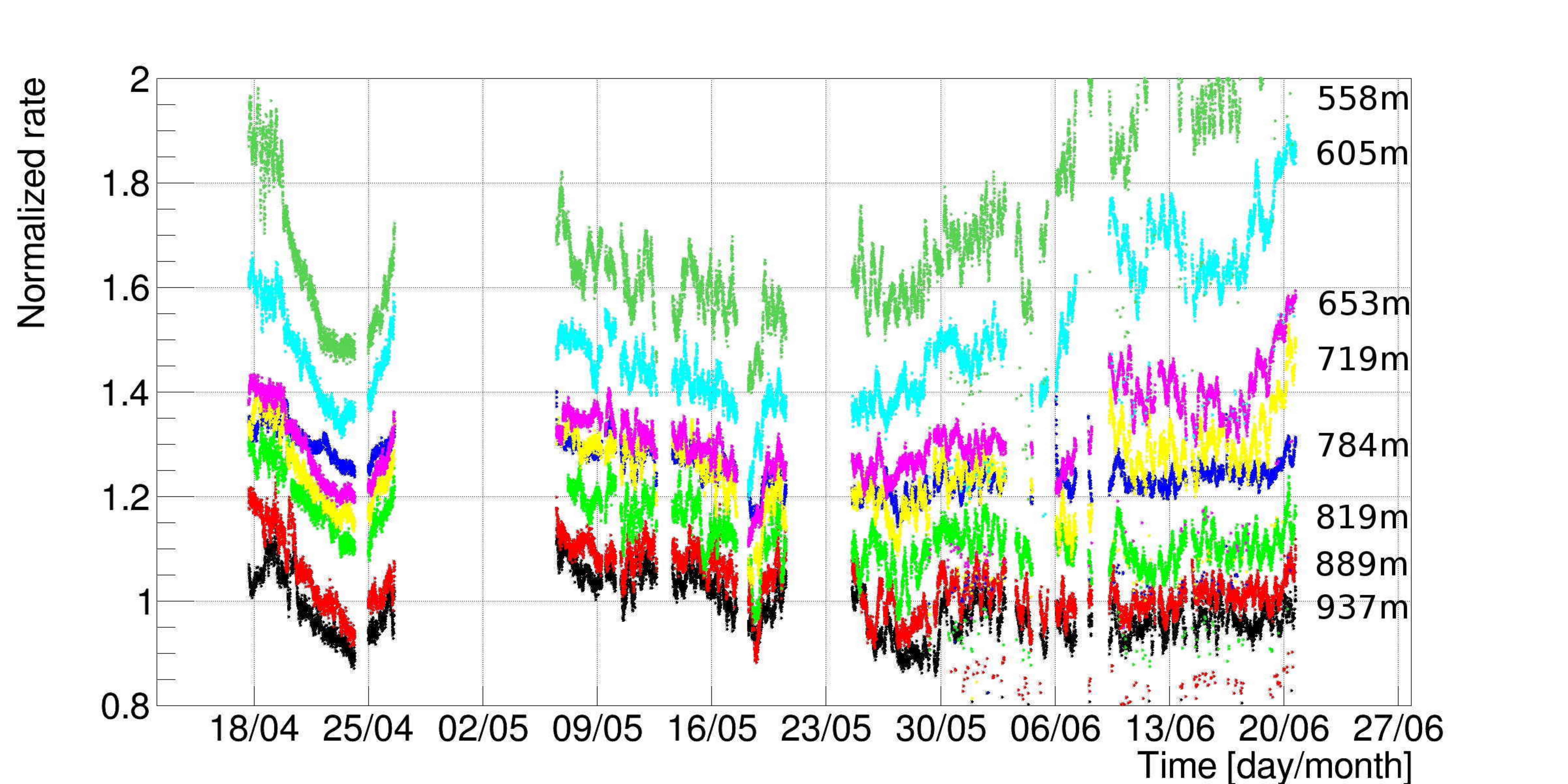}
  \caption{Normalized counting rates of OMs arranged at
 instrumentation string. For each curve the corresponding depth is
 shown.
}
  \label{fig12}
% \end{figure*}
 \end{figure}
%%%%%%%%%%%%%%%%%%%%%%%%%%%%
%%%%%%%%%%%%%%%%%%%%%%%%%%%%%%%%%%%%%%%%%%%%%%%%%%
\section{Instrumentation string}
\label{s7}
 The sketch of the instrumentation string is shown
 in figure 7. Instrumentation string is located at about
 of 100 m apart from the measuring strings with
 OMs. It comprises the calibration laser source, eight
 optical modules, as well as 10 acoustic sensors of
 positioning system.

 The calibration laser source \cite{PTE56,Yad.phys.69} is located at
 1215 m depth and is used for time synchronization
 between OMs on different strings. High intensity of
 laser source (up to $6\times10^{13}$ photons/pulse) allows
 illumination of OMs at distances more than 200 m
 from the source.

 Acoustic sensors are arranged along the
 instrumentation string starting from the 50 m depth
 to the bottom of string and perform monitoring of a
 string displacements at different depths caused by
 deep or/and surface water currents.

 As it was discussed earlier the optical modules of
 GVD design with 10386 OMs will be located at
 depths of 600 m to 1300 m below the surface, while
 OMs of the demonstration cluster are located at
 depths of 950 m to 1300 m. Eight optical modules
 housing R8055 or XP1807 PMTs are arranged at the
 depths from 600 m to 900 m on instrumentation
 string and aim at monitoring of a light background at
 these depths. Earlier these OMs were operated as the
 parts of engineering arrays during 2011-2012. In
 figure 12 the normalized counting rates of OMs
 arranged on the instrumentation string are shown.
 The rates are normalized on averaged counting rates
 recorded by same OMs during 2012 at the depths
 below 1200 m, where background slightly depends
 on the depth. As one can see from figure 12
 characteristic temporal behavior of background does
 not changed with depth, while intensity of light
 background increases with depth decreasing and
 becomes at 600 m depth twice larger comparing to
 intensity below 900 m. Preliminary results obtained
 in April--June 2013 indicate an opportunity to exploit
 the water volume below 600 m depth as
 instrumented volume of a neutrino telescope.
 However, the long-term monitoring is required to
 study a light background behavior during annual
 exposition of array at interested depths.

%%%%%%%%%%%%%%%%%%%%%%%%%%%%%%%%%%%%%%%%%%%%%%%%%%%%
\section{Conclusion}
\label{s8}
 The construction of a km$^3$-scale neutrino
 telescope -- the Gigaton Volume Detector in Lake
 Baikal -- is the central goal of the Baikal
 collaboration. During the R\&D phase of the GVD
 project in 2008-2010 years the basic elements of
 GVD -- new optical modules, FADC readout units,
 underwater communications and trigger systems --
 have been developed, produced and tested in situ by
 long-term operating prototype strings in Lake
 Baikal. The Prototyping phase of the GVD project
 has been started in April 2011 with the deployment
 of a three string engineering array which comprised
 all basic elements and systems of the GVD-telescope 
in Lake Baikal and was connected to
 shore by electro-optical cable. In April 2012 the
 first GVD-string with 24 OMs combined in two
 sections has been deployed as a part of three string
 engineering array. The first stage of GVD-cluster
 which comprises three strings has been deployed in
 2013. Deployment of the first demonstration GVD-cluster 
consisting of 8 strings is expected in 2015.

%%%%%%%%%%%%%%%%%%%%%%%%%%%%%%%%%%%%%%%%%%%%%%%%%%%%%

\section*{Acknowledgments}
 This work was supported by the Russian Ministry
 of Education and Science
 (agreement
 \#14.837.21.0785, \#8706, \#14.518.11.7074), by the
 Russian Found for Basic Research (grants 11-02-
 00983, 13-02-10012, 13-02-12221,
 13-02-00214, 13-02-10002).


%\bibliographystyle{elsarticle-num}



\end{document}